\documentclass[aps,prl,showpacs,amssymb,superscriptaddress,twocolumn]{revtex4-1}
\usepackage{graphicx}
\usepackage{appendix}
\usepackage{color}
\usepackage[caption=false]{subfig}
\usepackage{epstopdf} 
\DeclareGraphicsExtensions{.pdf,.eps,.png,.jpg,.mps}
\usepackage[colorlinks=true,citecolor=blue,urlcolor=blue,linkcolor=blue]{hyperref}
\usepackage{bm,amsmath,bbold,mathtools}
\usepackage{dcolumn}
\newcommand{\bra}[1]{\left\langle #1 \right|}
\newcommand{\ket}[1]{\left|#1\right\rangle}

\def\BEq{\begin{equation}}
\def\EEq{\end{equation}}
\def\BEqA{\begin{eqnarray}}
\def\EEqA{\end{eqnarray}}
\def\BW{\begin{widetext}}
\def\EW{\end{widetext}}

\begin{document}

\title{Pulse sequences for suppressing leakage in single-qubit gate operations}

\author{Joydip Ghosh}
\email{jghosh3@wisc.edu}
\affiliation{Department of Physics, University of Wisconsin-Madison, Madison, Wisconsin 53706, USA}
\author{S. N. Coppersmith}
\email{snc@physics.wisc.edu}
\affiliation{Department of Physics, University of Wisconsin-Madison, Madison, Wisconsin 53706, USA}
\author{Mark Friesen}
\email{friesen@physics.wisc.edu}
\affiliation{Department of Physics, University of Wisconsin-Madison, Madison, Wisconsin 53706, USA}

\date{\today}

\begin{abstract}
Many realizations of solid-state qubits involve couplings to leakage states lying outside the computational subspace, posing a threat to high-fidelity quantum gate operations. 
Mitigating leakage errors is especially challenging when the coupling strength is unknown, e.g., when it is caused by noise. 
Here we show that simple pulse sequences can be used to strongly suppress leakage errors for a qubit embedded in a three-level system.
As an example, we apply our scheme to the recently proposed charge quadrupole (CQ) qubit for  quantum dots. 
These results provide a solution to a key challenge for fault-tolerant quantum computing with solid-state elements.
\end{abstract}


\maketitle

Recent advances in semiconducting quantum dots make them promising candidates for universal quantum 
computing~\cite{PhysRevA.57.120,Elzerman2004,Petta2005,Hanson2007,Shulman2014,Veldhorst2015}.
However, performing gate operations with high enough fidelity to support fault-tolerant error correction remains a key challenge~\cite{Taylor2005}. Two ways that a qubit can fail to have high fidelity are (i) the qubit could decay or dephase, and (ii) quantum information could leak out of the qubit's logical subspace into other quantum states in the physical system~\cite{PhysRevA.71.052301,PhysRevLett.103.110501,West2012,PhysRevA.88.062329,PhysRevB.88.161303}. While several recent proposals in quantum dots have focused on suppressing dephasing from environmental noise~\cite{Kim2014,Eng2015}, relatively little effort has gone into suppressing leakage~\cite{PhysRevB.88.161303}.

Several approaches for reducing leakage errors have been developed for superconducting qubits, including analytic pulse shaping~\cite{PhysRevA.87.022309,PhysRevLett.103.110501} and optimal quantum control~\cite{0953-2048-27-1-014001,PhysRevLett.114.200502}.
{It is also known that leakage errors are, in principle, suppressible for a system-bath model with a composite sequence of a large number of pulses~\cite{PhysRevA.71.052301}.}
However, suppressing leakage errors below the fault-tolerant threshold under experimental conditions for quantum dot qubits is challenging, because of the need to apply smoothly varying {short} control pulses~\cite{Barnes2015} {within a time much less than the coherence times of the system}, and also because the qubits experience fluctuations that vary in time and strength~\cite{PhysRevB.94.045435}. 
For semiconducting systems, these constraints preclude using analytically derived pulse shapes, since these require knowing the noise strength as an input parameter~\cite{PhysRevLett.103.110501}, or quantum control strategies that rely on optimizing an unrealistically large number of control parameters~\cite{PhysRevLett.114.200502,0953-2048-27-1-014001}.

In this work, we develop a simple and experimentally feasible protocol based on composite pulses that suppresses leakage errors in a three-level quantum system where the leakage state is coupled to one of the logical states with an unknown but static coupling strength (referred to as the quasistatic noise approximation~\cite{PhysRevB.94.045435}). Whereas the leakage error scales quadratically with noise amplitude in conventional pulsed-gate schemes for this model, our protocol significantly improves the gate fidelities by eliminating computational errors in the logical subspace up to fourth order  in the noise amplitude
and leakage errors up to sixth order.

In a basis comprised of two logical states and one leakage state, the model Hamiltonian is given by
\BEq
\label{eq:ModelHamiltonian}
H=H_{\rm z} + H_{\rm x} + H_{\rm leak},
\EEq
with
\BEqA
&&H_{\rm z}=\frac{\epsilon_{\rm q}}{2}\left(\begin{array}{ccc}
1 & 0 & 0 \\
0 & -1 & 0 \\
0 & 0 & -\zeta \end{array}\right), \;\;\;
H_{\rm x}=g\left(\begin{array}{ccc}
0 & 1 & 0 \\
1 & 0 & 0 \\
0 & 0 & 0 \end{array}\right) \;\; {\rm and} \nonumber\\
&&H_{\rm leak}=\xi\left(\begin{array}{ccc}
0 & 0 & 0 \\
0 & 0 & 1 \\
0 & 1 & 0 \end{array}\right) \nonumber,
\EEqA
where $\epsilon_{\rm q}$ and $g$ are the independent control parameters for rotations about the $z$- and $x$-axes of the Bloch sphere  in the logical subspace, $\xi$ denotes the unknown coupling between the leakage state and one of the logical states, and $\zeta$ denotes the (scaled) leakage state energy in the absence of coupling.
While the Hamiltonian (\ref{eq:ModelHamiltonian}) provides a very general description of a 2-level system coupled to a leakage state~\cite{PhysRevA.88.052330,PhysRevLett.103.110501},
we focus here on the semiconducting charge quadrupole (CQ) qubit~\cite{Friesen:2016aa}, for which the logical states have different charge distributions but the same center of mass, yielding a qubit that is inherently protected from the predominant type of noise in this system:  uniform electric field fluctuations. 

The CQ qubit is formed in three adjacent semiconducting quantum dots sharing a single electron~\cite{Friesen:2016aa}. In the localized charge basis $\{\ket{100},\ket{010},\ket{001}\}$, where the basis states denote the electron being in the $1^{\rm st}$, $2^{\rm nd}$ or the $3^{\rm rd}$ dot respectively, the Hamiltonian is given by
\BEq
H_{\rm CQ}=\left(\begin{array}{ccc}
\epsilon_{\rm d} & t_{\rm A} & 0 \\
t_{\rm A} & \epsilon_{\rm q} & t_{\rm B} \\
0 & t_{\rm B} & -\epsilon_{\rm d} \end{array}\right)+\frac{U_{1}+U_{3}}{2},
\EEq
where $U_{1,2,3}$ are the on-site potentials for the three dots, $t_{A,B}$ are tunnel couplings between adjacent dots, and $\epsilon_{\rm d}=(U_{1}-U_{3})/2$ and $\epsilon_{\rm q}=U_{2}-(U_{1}+U_{3})/2$ denote the dipolar and quadrupolar detuning parameters, respectively. We now define a new set of basis states~\cite{Friesen:2016aa}
\BEq
\ket{C}=\ket{010},
\ket{E}=\frac{\ket{100}+\ket{001}}{\sqrt{2}},
\ket{L}=\frac{\ket{100}-\ket{001}}{\sqrt{2}},
\label{eq:basisDef}
\EEq
where $\ket{C}$ and $\ket{E}$ correspond to logical states and $\ket{L}$ denotes the leakage state. In this basis, the Hamiltonian becomes
\BEq
\label{eq:Hcqcel}
\tilde{H}_{\rm CQ}=\left(\begin{array}{ccc}
\frac{\epsilon_{\rm q}}{2} & \frac{t_{\rm A}+t_{\rm B}}{\sqrt{2}} & \frac{t_{\rm A}-t_{\rm B}}{\sqrt{2}} \\
\frac{t_{\rm A}+t_{\rm B}}{\sqrt{2}} & -\frac{\epsilon_{\rm q}}{2} & \epsilon_{\rm d} \\
\frac{t_{\rm A}-t_{\rm B}}{\sqrt{2}} & \epsilon_{\rm d} & -\frac{\epsilon_{\rm q}}{2} \end{array}\right) ,
\EEq
where we have neglected a term proportional to the identity. Tunnel couplings and detuning parameters can be independently tuned in quantum dots~\cite{PhysRevLett.116.110402,PhysRevLett.116.116801}, and we set them here to $t_{\rm A}=t_{\rm B}$ and $\epsilon_{\rm d}=0$. In this case, it can be shown that $\tilde{H}_{\rm CQ}$ forms a decoherence-free subspace (DFS) with respect to uniform electric field fluctuations~\cite{Friesen:2016aa}. 
Fluctuations of the dipolar detuning parameter $\epsilon_{\rm d}$ break this DFS by coupling the states $\ket{E}$ and $\ket{L}$. 

\begin{figure}[htb]
\centering
\includegraphics[angle=0,width=3.3in]{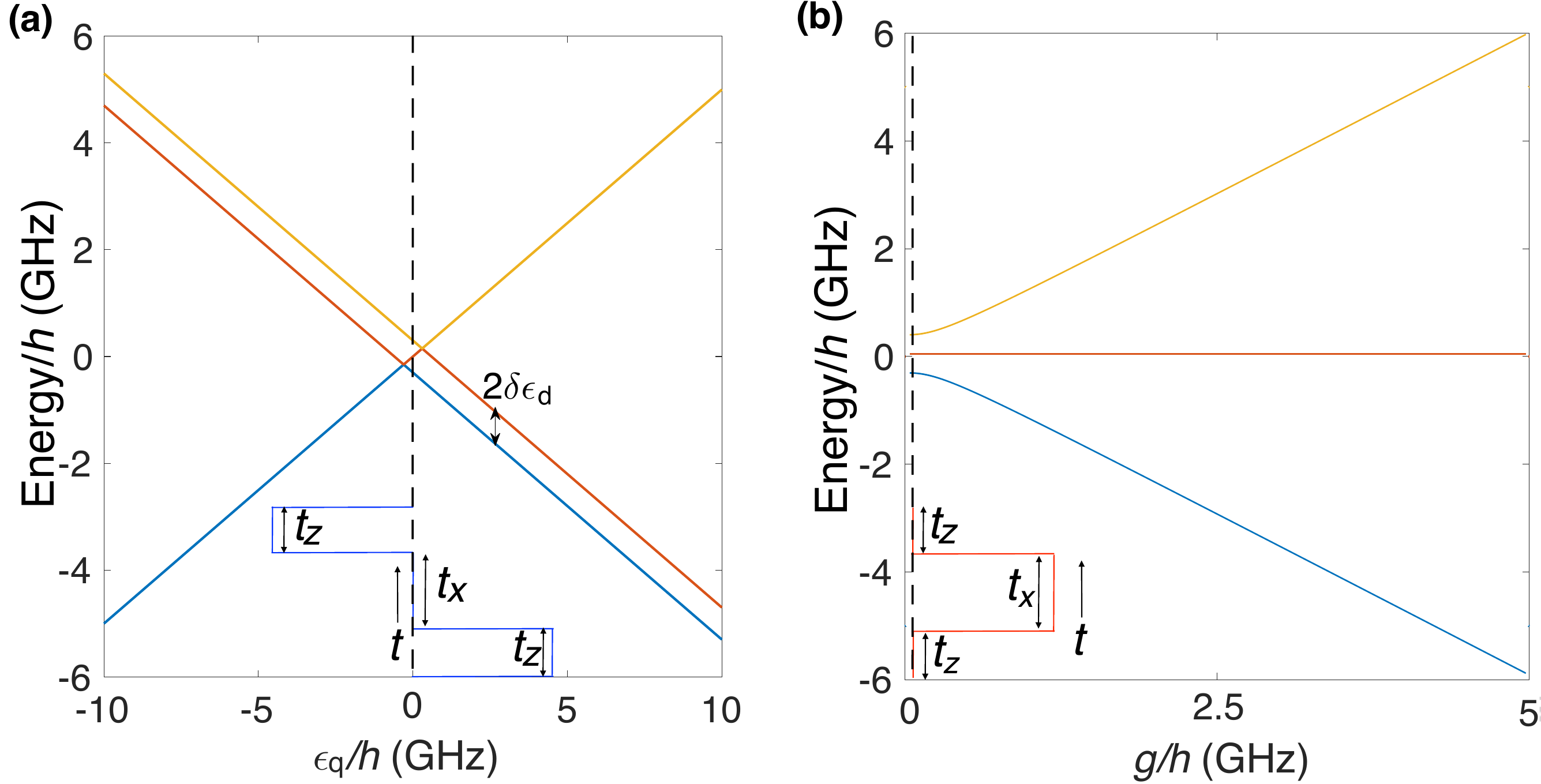}
\caption{Energy eigenvalues of the CQ qubit, as functions of (a) $\epsilon_{\rm q}$ (with $g/h=0$ and $\delta\epsilon_{\rm d}/h=0.3$ GHz) and (b) $g$ (with $\epsilon_{\rm q}/h=0$ and $\delta\epsilon_{\rm d}/h=0.3$ GHz). The vertical dashed lines denote the idle point ($\epsilon_{\rm q}=g=0$) for the CQ qubit. The insets (which use the same time axis) show the bang-bang pulse profiles for $\epsilon_{\rm q}$ and $g$ that implement the $\mathcal{R}_{\rm zxz}$ composite gate [see Eq.~(\ref{eq:RzxzDefinition})].
These pulses generate $z$-rotations in (a), and $x$-rotations in (b).}
\label{fig:energyLevels}
\end{figure}

Under the conditions described above, $\tilde{H}_{\rm CQ}$ reduces to Eq.~(\ref{eq:ModelHamiltonian}), where $\zeta=1$, $\xi=\epsilon_{\rm d}$, $g=(t_{\rm A}+t_{\rm B})/\sqrt{2}$. 
{For the remainder of this paper, we will adopt Eq.~(\ref{eq:Hcqcel}) with $t_{\rm A}=t_{\rm B}$ as our model Hamiltonian. We restrict our analysis to the special case $\zeta=1$, as appropriate for CQ qubits, although it is also possible to generalize our approach to the case $\zeta \neq 1$ by considering longer pulse sequences that are more challenging to implement, experimentally.}
Figure~\ref{fig:energyLevels} shows the energy levels of $\tilde{H}_{\rm CQ}$ for typical device parameters, as a function of $\epsilon_{\rm q}$ and $g$.
Here, the vertical dashed lines denote the idle point for qubit gate operations. 
A rotation about the $z$-axis of the Bloch sphere can now be performed by pulsing $\epsilon_{\rm q}$ (with $g=0$) from the idle point to a region with $|\epsilon_{\rm q}| \gg 0$, as shown in Fig.~\ref{fig:energyLevels}(a).
Here the large energy gap between logical states $\ket{C}$ and $\ket{E}$ generates the phase required for the rotation. 
Similarly, a rotation about the $x$-axis of the Bloch sphere is performed by pulsing $g$ (with $\epsilon_{\rm q}=0$) from the idle point, as shown in Fig.~\ref{fig:energyLevels}(b).

The predominant noise in this system arises from the motion of trapped charge in the semiconductor or dielectric materials~\cite{Dial2013,Wu19082014,BrandurPreprint}.
We model it as  $\epsilon_{\rm q}=\bar{\epsilon}_{\rm q}+\delta\epsilon_{\rm q}$ and $\epsilon_{\rm d}=\bar{\epsilon}_{\rm d}+\delta\epsilon_{\rm d}$, where $\bar{\epsilon}_{\rm q,d}$ are the average quadrupolar and dipolar detuning control parameters (here, $\bar \epsilon_\text{d}=0$), and $\delta\epsilon_{\rm q,d}$ are the corresponding fluctuations. 
In \cite{Friesen:2016aa} it is argued that, while $\delta\epsilon_{\rm d}\gg \delta\epsilon_{\rm q}$, the effects of $\delta\epsilon_{\rm d}$ are largely suppressed by the DFS.
Here, we focus on the residual effects of $\delta\epsilon_{\rm d}$, which primariliy cause leakage.  
Experimentally, it is known that the noise spectrum of $\delta\epsilon_{\rm d}$ is dominated by low-frequency fluctuations, which are slow compared to gate operations~\cite{Dial2013,Kawakami2016}. 
We therefore assume that while $\delta\epsilon_{\rm d}$ is variable, its value remains constant during a given gate operation~\cite{PhysRevB.94.045435}. {In this work, we neglect the less dominant sources of noise, such as high-frequency components and control jitters for voltage and timing.}

We now show how leakage due to $\delta\epsilon_{\rm d}$ can be suppressed for arbitrary single-qubit rotations. We first consider the ``bang-bang" limit, in which the control pulses switch instantaneously between two values, and we analytically determine the probabilities for leakage and computational errors.  Later, we will relax the bang-bang constraint and numerically compute the fidelities of single-qubit rotations with smoothly varying pulse profiles, which are more realistic for controlling semiconducting qubits.

The unitary operators for noisy $z$- and $x$-rotations are
\BEqA
U_{\rm z}(\epsilon_{\rm q},\delta\epsilon_{\rm d},\varphi)&=&e^{-i\left[H_{\rm z}(\epsilon_{\rm q})+H_{\rm leak}(\delta\epsilon_{\rm d})\right]\varphi/\epsilon_\text{q}} \;\; {\rm and} \nonumber \\
U_{\rm x}(g,\delta\epsilon_{\rm d},\theta)&=&e^{-i\left[H_{\rm x}(g)+H_{\rm leak}(\delta\epsilon_{\rm d})\right]\theta/2g} ,
\label{eq:UzUx}
\EEqA
for arbitrary rotation angles $\varphi$ and  $\theta$.
For bang-bang gates, these angles are related to the corresponding gate times as $\varphi=t_z(\epsilon_\text{q}/\hbar)$ and $\theta=t_x(2g/\hbar)$.
Hence, $\varphi$ must have the same sign as $\epsilon_\text{q}$, and $\theta$ must be positive, since $g> 0$  for most solid-state devices, including quantum dots.
Our approach will be to compose $U_\text{z}$ and $U_\text{x}$ gates to obtain arbitrary single-qubit rotations in the logical subspace as a function of $\epsilon_\text{q}$, $g$, $\varphi$, and $\theta$, and then determine the conditions that should be imposed on these parameters to suppress the leading order leakage terms in the composed gates.

If $z$- and $x$-rotations are performed using ``bare" $U_z$ and $U_x$ operations, i.e.\ with no gate sequence, it can be shown that leakage error, and hence the computational error in the logical subspace, both scale as $(\delta\epsilon_{\rm d}/g)^{2}$~\cite{Friesen:2016aa}. 
In \cite{Supp} we show that it is impossible to achieve better results by composing just two bare gates. 
We therefore focus here on compositions of three bare gates, of the form
\BEq
\label{eq:RzxzDefinition}
\mathcal{R}_{\rm zxz}(\theta,\varphi)=U_{\rm z}(\epsilon_{\rm q},\delta\epsilon_{\rm d},\frac{\varphi}{2})U_{\rm x}(g,\delta\epsilon_{\rm d},\theta)U_{\rm z}(-\epsilon_{\rm q},\delta\epsilon_{\rm d},\frac{-\varphi}{2}).
\EEq
A schematic diagram of a bang-bang pulse sequence for this $\mathcal{R}_{\rm zxz}$ operation is shown in the insets of Fig.~\ref{fig:energyLevels}.

We now expand $\mathcal{R}_{\rm zxz}$ about $\delta\epsilon_{\rm d}=0$ and note that the first order terms of $\delta\epsilon_{\rm d}$ vanish from the Taylor expansion if we set~\cite{Supp}
\BEq
\epsilon_{\rm q}={-\frac{g\varphi}{2}\cot\left(\frac{\theta}{4}\right)}.
\label{eq:zxzCondition}
\EEq
This condition, along with the conditions on $\varphi$, $\theta$, and $g$, are all satisfied when $2\pi < \theta < 4\pi$, which means that $U_x$ rotations are between 1 and 2 cycles around the Bloch sphere.

We note that $\mathcal{R}_{\rm zxz}(\theta,\varphi)$ corresponds to a rotation of angle $\theta$ about an arbitrary axis $\hat{n}=\cos(\varphi/2)\hat{x}+\sin(\varphi/2)\hat{y}$ in the $x$-$y$ plane of the logical subspace (modulo an irrelevant overall phase). 
To estimate analytically how the computational and leakage errors depend on the noise $\delta\epsilon_{\rm d}/g$, we evaluate the higher order terms in the Taylor expansion of $\mathcal{R}_{\rm zxz}(\theta,\varphi)$~\cite{Supp}, observing that the error probability within the logical subspace $|\mathcal{R}_{\rm zxz}({\delta\epsilon_{\rm d} \neq 0})-\mathcal{R}_{\rm zxz}({\delta\epsilon_{\rm d}=0})|^2$ scales as $(\delta\epsilon_{\rm d}/g)^{4}$, while the probability of leakage errors $P_{LC,LE}=\left|\bra{L}\mathcal{R}_{\rm zxz}\ket{C,E}\right|^{2}$ scales as $(\delta\epsilon_{\rm d}/g)^{6}$. {Therefore, for a typical experimental value of $\delta\epsilon_{\rm d}/g \sim 10^{-1}$~\cite{BrandurPreprint}, the error in the computational subspace is $\sim 10^{-4}$ and $P_{LC,LE} \sim 10^{-6}$.} This represents a remarkable improvement over the bare gates, {for which the leakage error is of order $10^{-2}$ when $\delta\epsilon_{\rm d}/g \sim 10^{-1}$ (For a detailed comparison of leakage errors in simple and composite pulse sequences, see~\cite{Supp})}.

Finally, we construct a completely general and arbitrary rotation on the Bloch sphere.
Since the effective rotation axis for an $\mathcal{R}_{\rm zxz}$ gate lies anywhere in the $x$-$y$ plane, an arbitrary rotation requires just two steps~\cite{2013arXiv1303.0297S}: $\mathcal{R}_{\rm zxz}(\theta_2,\varphi_2)\mathcal{R}_{\rm zxz}(\theta_1,\varphi_1)$, comprising 6 bare gates.  
Although there are penalties associated with longer gate sequences, such as reduced gate speeds and accumulation of errors,
the sequences described here improve the overall scaling of errors with respect to the noise amplitude.
Hence, they are always beneficial if $\delta\epsilon_\text{d}$ is small enough. 

While an $\mathcal{R}_{\rm zxz}$ gate produces arbitrary rotations in the $x$-$y$ plane, Eq.~(\ref{eq:zxzCondition}) indicates that the control parameter $g$ ($\epsilon_\text{q}$) diverges in the limit $\theta\rightarrow 2\pi$ ($4\pi$). This implies that an identity gate cannot be implemented using a single $\mathcal{R}_{\rm zxz}$ operation. 
A high-fidelity identity gate is an essential ingredient for fault-tolerant quantum computing because most qubits remain idle during most of the quantum error-correction cycle.  
We cannot use the ``idle gate," corresponding to $\epsilon_\text{q}=g=0$, as an identity gate because $\delta\epsilon_\text{d}\neq 0$ causes errors.
It is possible to obtain an identity gate from the sequence $[\mathcal{R}_{\rm zxz}(\pi,\varphi)]^2$, which comprises 6 bare gates; however it is interesting to ask whether shorter identity sequences exist.

We now present a viable identity sequence, $\mathcal{R}_{I}$, that uses just 4 bare gates:
Following the same procedure used to construct $\mathcal{R}_{\rm zxz}$, we consider the sequence 
\BEq
\label{eq:Identity}
\mathcal{R}_{I}=\left[U_{\rm z}(\epsilon_{\rm q},\delta\epsilon_{\rm d},2\pi)U_{\rm x}(g,\delta\epsilon_{\rm d},2\pi)\right]^2.
\EEq
$\mathcal{R}_{I}$ is clearly an identity operation when $\delta\epsilon_{\rm d}=0$. 
To address the case $\delta\epsilon_{\rm d} \neq 0$, we perform a Taylor expansion of $\mathcal{R}_{I}$ about $\delta\epsilon_{\rm d}=0$ \cite{Supp}, noting that the first-order term in $\delta\epsilon_{\rm d}$ vanishes due to the special form of the sequence, even without imposing special constraints on $\epsilon_q$ and $g$. The Taylor expansion also indicates that the probability of computational errors, given by $\left|\bra{C}\mathcal{R}_{I}\ket{E}\right|^{2} = \left|\bra{E}\mathcal{R}_{I}\ket{C}\right|^{2}$, scales as $(\delta\epsilon_{\rm d}/g)^{4}$, while the probability of leakage errors, given by $\left|\bra{L}\mathcal{R}_{I}\ket{C}\right|^{2} = \left|\bra{L}\mathcal{R}_{I}\ket{E}\right|^{2}$, scales as $(\delta\epsilon_{\rm d}/g)^{6}$.

To understand why our pulse sequences work, we now describe a method for simultaneously visualizing the evolution of the logical and leakage states on a pair of coupled Bloch spheres. An arbitrary state $\ket{\Psi}$ of a three-level quantum system can be expressed in the $\left\{\ket{C}, \ket{E}, \ket{L}\right\}$ basis using four angle variables:
\BEq
\ket{\Psi}=\begin{pmatrix}
\cos(\vartheta/2)\cos(\chi/2) \\
\cos(\vartheta/2)\sin(\chi/2)e^{i\varrho} \\
\sin(\vartheta/2)e^{i\varsigma} \end{pmatrix} .
\EEq
When $\vartheta=0$, $\ket{\Psi}$ maps onto $\ket{\psi_{g}}=\cos(\chi/2)\ket{C}+\sin(\chi/2)e^{i\varrho}\ket{E}$, comprising a two-level system. 
Similarly, when $\chi=0$, $\ket{\Psi}$ maps onto $\ket{\psi_{r}}=\cos(\vartheta/2)\ket{C}+\sin(\vartheta/2)e^{i\varsigma}\ket{L}$, comprising a different two-level system. 
The full mapping $\ket{\Psi}\rightarrow \{\ket{\psi_g},\ket{\psi_r}\}$ is bijective, and the states $\ket{\psi_{g}}$ and $\ket{\psi_{r}}$ may alternatively be represented as unit vectors on a pair of Bloch spheres that we label `green' and `red,' respectively.
The angles $\vartheta$, $\chi$, $\varrho$, and $\varsigma$ here correspond to the usual polar and azimuthal angles of the respective Bloch spheres.

In the absence of leakage, the state vector of the red sphere points to the north pole and the state vector of the green sphere describes the logical qubit in the usual way. 
When leakage occurs, the state vector of the red sphere deviates from north, with its latitude describing the amplitude of the leakage state, and its azimuth describing the relative phase of $\ket{L}$ with respect to $\ket{C}$.
The green sphere still describes the logical qubit, with its state vector renormalized to lie on the unit sphere.
For quantum systems with multiple leakage states, we could extend this geometrical representation to include multiple Bloch spheres, with one sphere describing the logical states, and one additional sphere for each leakage state.
Below, we restrict our analysis to a single leakage state and compute the time-dependent trajectories of the green and red Bloch spheres for the $\mathcal{R}_{\rm zxz}$ pulse sequence. 

\begin{figure}[htb]
\centering
\label{fig:Evolution}
\includegraphics[angle=0,width=3.2in]{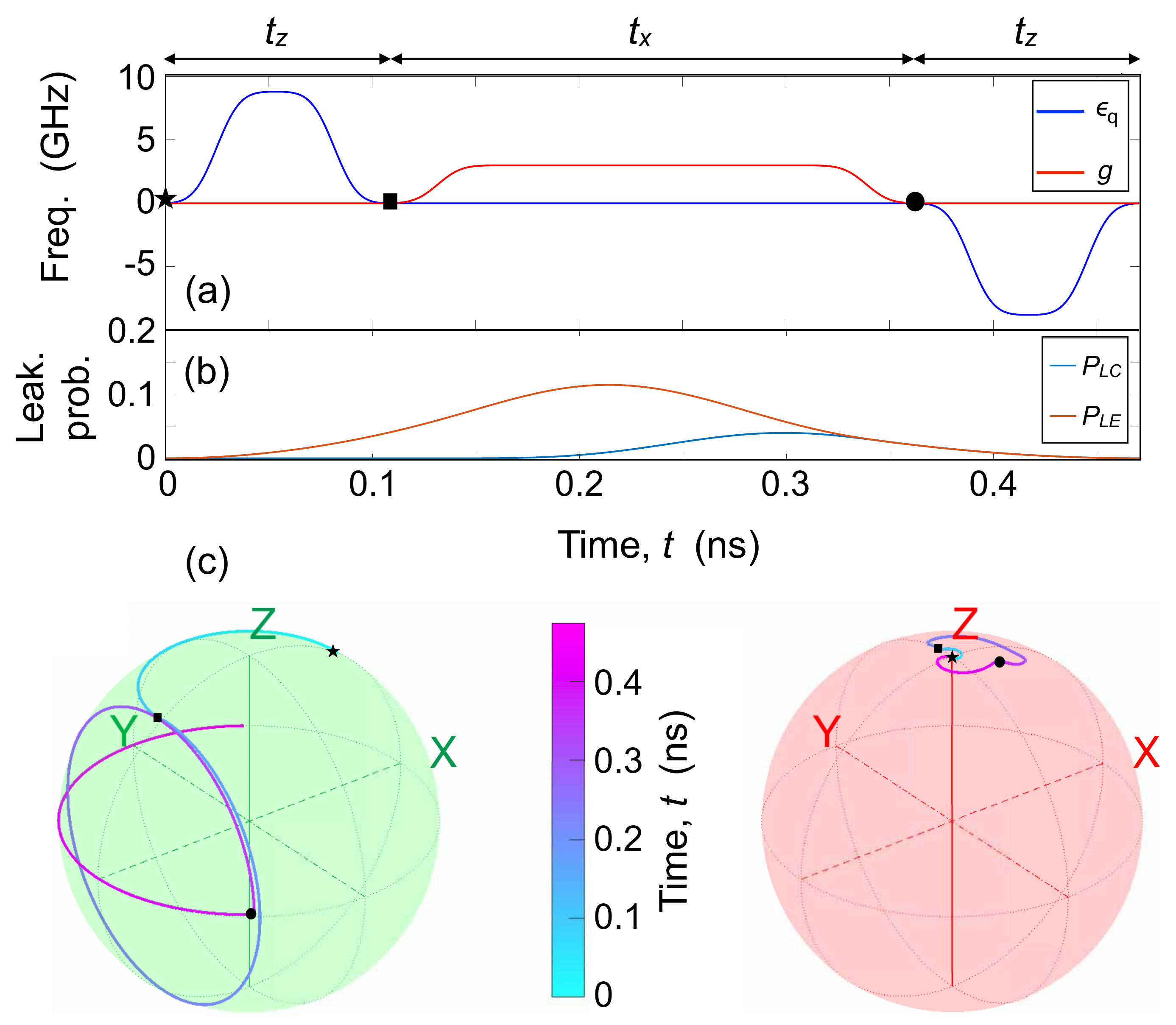}
\caption{A numerically optimized gate operation.
(a) The pulse sequence for an $X_{-\pi/2}$ gate, assuming rounded pulse profiles with $50$~ps rise times.
Here, the optimization parameters are the gate times $t_z$ and $t_x$, and the peak value of $\epsilon_\text{q}$.
The peak value of $g/h=3$~GHz is fixed.
(b)  Leakage probabilities as a function of time.
The final leakage error is of order $10^{-10}$.
(c) Time evolution of the logical and leakage states on green and red ``Bloch spheres" respectively, {for the smooth pulse shown in (a).}
Here, the color of the state vectors indicates the time.
The transitions between $z$-, $x$-, and $z$-rotations are indicated by symbols, as consistent with (a).}
\label{fig:Evolution}
\end{figure}

Our analytical results for pulse sequences were previously obtained in the bang-bang limit.
However in charge qubit experiments, the combination of high-frequency filtering and short gate times ($\sim 0.1$-1~ns) means that pulses will experience significant rounding.
To account for this, we now consider pulse sequences with smooth profiles and finite rise times. 
Under such conditions, constraints on the pulse shape, such as Eq.~(\ref{eq:zxzCondition}), must be modified.
To do this, we keep our previous sequences, but vary the control parameters and gate times.
The parameters are chosen to maximize the process fidelity, defined as~\cite{Pedersen2007,PhysRevA.81.052340}
\BEq
F=\frac{{\rm Tr}({\mathcal U}{\mathcal U}^{\dagger})+|{\rm Tr}(U_{\rm target}^{\dagger}{\mathcal U})|^{2}}{d(d+1)} .
\label{eq:FidelityDef}
\EEq
{Here, $U_\text{target}$ is the desired gate operation in the 2D logical subspace, $\mathcal U$ is the unitary operation obtained from simulations and projected onto the 2D logical subspace, and $d=2$.}
Such projections do not generally yield unitary operators; however Eq.~(\ref{eq:FidelityDef}) takes this into account, and captures both computational and leakage errors.

A typical result from our gate optimization procedure is shown in Fig.~\ref{fig:Evolution}.
Here, we consider an effective  $X_{-\pi/2}$ rotation, obtained from the 3-step sequence defined in Fig.~\ref{fig:Evolution}(a), which is analogous to the bang-bang sequence $\mathcal{R}_{\rm zxz}(2\pi+\pi/2,2\pi)$. 
We adopt several fixed device parameters, including error-function pulse profiles with rise times of 50~ps~\cite{Supp}, a peak tunnel coupling of $g/h=3$~GHz, and a typical experimental noise value of $\delta \epsilon_{d}/h = 0.3$~GHz~\cite{BrandurPreprint}.
We then numerically optimize the gate times $t_z$ and $t_x$ defined in the figure, and the peak value of $\epsilon_q$, to maximize the fidelity.
Figures~\ref{fig:Evolution}(b) and \ref{fig:Evolution}(c) show the time evolution of the resulting gate for the initial state $\cos(\pi/10)\ket{C}+\sin(\pi/10)\ket{E}$.
Although there is significant leakage in the middle of the operation, the final leakage probability is strongly suppressed.
From Eq.~(\ref{eq:FidelityDef}), the final infidelity is found to be $1-F=2.46\times10^{-8}$, while the total leakage errors corresponding to $\ket{C}$ and $\ket{E}$ states are $1.5\times10^{-10}$ and $1.2\times10^{-10}$ respectively. {These results are obtained for the smooth pulse sequence shown in Fig.~\ref{fig:Evolution}(a), and represent orders of magnitude of improvement in the infidelity compared to the bang-bang pulse under the constraint~(\ref{eq:zxzCondition}). This improvement can be attributed to the numerical optimization procedure. Indeed, by relaxing constraint~(\ref{eq:zxzCondition}), while retaining the three-pulse construction, the dynamics can be modified to cancel out higher order noise effects, for either bang-bang or smooth pulses.}

The benefit of the two-Bloch-sphere representation is apparent in Fig.~\ref{fig:Evolution}(c), where it provides a qualitative understanding of how our pulse sequence suppresses leakage errors.
Starting from the north pole of the red sphere, the leakage probability grows slowly during the first $z$-rotation.
As it grows, the leakage state accumulates phase, causing the state vector to spiral outward and downward.
The special constraint on the control parameters, analogous to Eq.~(\ref{eq:zxzCondition}), causes the leakage phase to be inverted during the $x$-rotation, similar to a Hahn echo.
The final $z$-rotation then yields a time evolution that is opposite of the initial evolution (a reverse spiral), which eventually passes very close to the north pole.

In conclusion, we have developed special pulse sequences for a three-level system, comprised of two logical states and a leakage state, which suppress the leakage probability and enable arbitrary, high-fidelity, single-qubit rotations.
We consider a general model, in which only one of the logical states is coupled to the leakage state.
A crucial feature of this method is that it requires no knowledge of the coupling to the leakage state.
The method is perfectly suited for the quantum dot charge quadrupole (CQ) qubit, whose leakage coupling is caused by charge noise.
Taking the CQ as our model system we have shown that computational errors in the gate operation arise at fourth order in the noise amplitude, while leakage errors arise at sixth order. This represents a substantial improvement over conventional pulse schemes, for which errors arise at second order in the noise amplitude. {While it is possible to design high-fidelity quantum gates for quantum dot qubits using optimal control schemes~\cite{0953-2048-27-1-014001,PhysRevLett.114.200502}, we emphasize that our approach yields high gate fidelities without knowing the amplitude of the noise, which sets it apart from those schemes. We also note that a Hahn echo sequence, which has the same level of complexity as our sequence, has recently been employed successfully to improve the fidelities of charge dipole qubits~\cite{Kim2015}. This suggests that our scheme could also yield high-fidelity gate operations in charge quadrupole qubits.}
To understand why our pulse sequence works, we have introduced an intuitive geometrical mapping of the three-level system, in which the logical and leakage states are represented on two coupled Bloch spheres. 
In this way, the cancellation of leakage errors appears similar to a Hahn echo.
Our proposal therefore provides an important step towards fault-tolerant quantum computation in quantum dots. 

We thank Mark Eriksson and Brandur Thorgrimsson for many helpful discussions. The authors would like to acknowledge support from the Vannevar Bush Faculty Fellowship program sponsored by the Basic Research Office of the Assistant Secretary of Defense for Research and Engineering and funded by the Office of Naval Research through grant N00014-15-1-0029. This work was also supported in part by Army Research Office Grant W911NF-12-0607 and National Science
Foundation (NSF) Grant PHY-1104660.


%

\onecolumngrid
\newpage
\renewcommand{\thesection}{S\arabic{section}}   
\renewcommand{\thetable}{S\arabic{table}}   
\renewcommand{\thefigure}{S\arabic{figure}}
\renewcommand{\theequation}{S\arabic{equation}}
\setcounter{equation}{0}

\section{Supplemental Material for ``Pulse sequences for suppressing leakage in single-qubit gate operations"}

In these Supplemental Materials we first show that two-pulse sequences cannot correct leakage errors, which leads us to study three-pulse sequences. We then provide details of the derivations of internal errors produced by our gate protocols, as well as the pulse shapes used in our simulations. Finally, we present a comparison of gate process fidelities between simple and composite pulse sequences.

\subsection{Demonstration that a composition of two pulses cannot correct leakage errors}
As mentioned in the main text, the leakage error for bare $x$- and $z$-rotations implemented in a charge quadrupole qubit scales as $(\delta\epsilon_{\rm d}/g)^2$. Here we show that composing two such pulses is also insufficient for suppressing the leading-order leakage errors. 

We consider the following two composed operations,
\BEqA
U_{\rm zx}(\epsilon_{\rm q},g,\delta\epsilon_{\rm d},\theta,\varphi)&=& U_{\rm z}(\epsilon_{\rm q},\delta\epsilon_{\rm d},\varphi)U_{\rm x}(g,\delta\epsilon_{\rm d},\theta) \;\;\; {\rm and} \nonumber \\
U_{\rm xz}(\epsilon_{\rm q},g,\delta\epsilon_{\rm d},\theta,\varphi)&=& U_{\rm x}(g,\delta\epsilon_{\rm d},\theta)U_{\rm z}(\epsilon_{\rm q},\delta\epsilon_{\rm d},\varphi),
\label{eq:UzUx}
\EEqA
where the bare gates are defined as
\BEqA
U_{\rm z}(\epsilon_{\rm q},\delta\epsilon_{\rm d},\varphi)&=&e^{-i\left[H_{\rm z}(\epsilon_{\rm q})+H_{\rm leak}(\delta\epsilon_{\rm d})\right]\frac{\varphi}{\epsilon_{\rm q}}} \;\; {\rm and} \nonumber \\
U_{\rm x}(g,\delta\epsilon_{\rm d},\theta)&=&e^{-i\left[H_{\rm x}(g)+H_{\rm leak}(\delta\epsilon_{\rm d})\right]\frac{\theta}{2g}},
\label{eq:UzUx}
\EEqA
and the various parameters and expressions are defined the same as in the main text.

First we consider $U_{\rm zx}$.
Occupation of the leakage state can be expressed in terms of the matrix elements $\bra{L}U_{\rm zx}\ket{C}$ and $\bra{L}U_{\rm zx}\ket{E}$.
Expanding these two expressions about $\delta\epsilon_{\rm d}=0$ yields leading-order terms proportional to 
\BEq
\label{eq:UZXCond}
2\epsilon_{\rm q} \sin^2\left(\frac{\theta }{4}\right)+g \varphi  \sin\left(\frac{\theta }{2}\right) 
\hspace{.3in}\text{and}\hspace{.3in}
\epsilon_{\rm q} \sin \left(\frac{\theta }{2}\right)+g \varphi  \cos \left(\frac{\theta }{2}\right),
\EEq
respectively.
To suppress the leakage errors to leading order, we must simultaneously set both of these expressions to zero.
It is easy to show that the only solutions to this problem involve ``null" rotations, for which either $\epsilon_\text{q}=0$ or $g=0$.
However, these do not represent valid solutions, because $U_\text{zx}$ then reduces to one of the bare rotations, $U_\text{z}$ or $U_\text{x}$, which provide no leakage protection.

Applying the same considerations to $U_\text{xz}$ yields the constraints
\BEq
\label{eq:UXZCond}
\cos\left(\frac{\theta }{2}\right)-1=\epsilon_{\rm q} \sin\left(\frac{\theta }{2}\right)+g \varphi=0,
\EEq
which also do not allow viable solutions.
This proves that it is impossible to suppress the leading-order effects of leakage using compositions of just two bare rotations. 

{One can also attempt to compose two bare $z$- or $x$-rotations. However, it can be explicitly shown that leading order leakage error cannot be suppressed for either of these pulses under realistic conditions.} We therefore turn to compositions of three bare rotations.

\subsection{Analysis of errors for $\mathcal{R}_{\rm zxz}(\theta,\varphi)$}
\label{sec:Rzxz}
In this section we consider the three-pulse sequence $\mathcal{R}_{\rm zxz}(\theta,\varphi)$ and determine the requirements for achieving noise suppression in this case.

A Taylor expansion of $\mathcal{R}_{\rm zxz}(\theta,\varphi)$ about $\delta\epsilon_{\rm d}=0$ yields
{\small \BEq
\label{eq:RzxzExpanded}
\mathcal{R}_{\rm zxz}(\theta,\varphi)=
\left(\begin{array}{ccc}
\cos\left(\frac{\theta }{2}\right) & -ie^{-\frac{i \varphi}{2}} \sin \left(\frac{\theta }{2}\right) & 0 \\
-ie^{\frac{i \varphi}{2}} \sin \left(\frac{\theta }{2}\right) & \cos \left(\frac{\theta }{2}\right) & 0 \\
0 & 0 & 1 \\
\end{array}\right)-\left(\frac{g\varphi \cos\left(\frac{\theta}{4}\right)}{\epsilon_{\rm q}}+2\sin \left(\frac{\theta}{4}\right)\right)
\left(\begin{array}{ccc}
0 & 0 & e^{-\frac{i \varphi}{2}} \sin \left(\frac{\theta}{4}\right) \\
0 & 0 & i\cos \left(\frac{\theta}{4}\right) \\
e^{\frac{i \varphi}{2}} \sin \left(\frac{\theta}{4}\right) & i\cos \left(\frac{\theta}{4}\right) & 0 \\
\end{array}\right)\frac{\delta\epsilon_{\rm d}}{g} ,
\EEq}
up to $O[\delta\epsilon_\text{d}/g]$.
This $O[\delta\epsilon_\text{d}/g]$ term clearly causes leakage.

We can make the $O[\delta\epsilon_\text{d}/g]$ term vanish by choosing control parameters such that 
\BEq
\epsilon_{\rm q}={-\frac{g\varphi}{2}\cot\left(\frac{\theta}{4}\right)}.
\label{eq:SzxzCondition}
\EEq
When this condition is satisfied, the next-leading-order terms in the Taylor expansion become
\BEq
\label{eq:RzxzExpanded}
\mathcal{R}_{\rm zxz}(\theta,\varphi)=
\left(\begin{array}{ccc}
\cos\left(\frac{\theta }{2}\right) & -ie^{-\frac{i \varphi}{2}} \sin \left(\frac{\theta }{2}\right) & 0 \\
-ie^{\frac{i \varphi}{2}} \sin \left(\frac{\theta }{2}\right) & \cos \left(\frac{\theta }{2}\right) & 0 \\
0 & 0 & 1 \\
\end{array}\right) +
\left(\begin{array}{ccc}
\mathcal{A}(\theta) & i e^{-\frac{i\varphi}{2}}\mathcal{B}(\theta) & 0 \\
 i e^{\frac{i\varphi}{2}}\mathcal{B}(\theta) & \mathcal{A}(\theta) & 0 \\
0 & 0 & \mathcal{C}(\theta,\varphi) \\
\end{array}\right)\left(\frac{\delta\epsilon_{\rm d}}{g}\right)^{2}+\mathcal{O}\left(\frac{\delta\epsilon_{\rm d}^{3}}{g^3}\right),
\EEq
where
\begin{gather}
\mathcal{A}(\theta)=2\sin\left(\frac{\theta}{4}\right)\left[\sin\left(\frac{\theta}{4}\right)-\frac{\theta}{4}\cos\left(\frac{\theta}{4}\right)\right], \nonumber \\
\mathcal{B}(\theta)=\sin\left(\frac{\theta}{2}\right)-\frac{\theta}{4}\cos\left(\frac{\theta}{2}\right)-\tan\left(\frac{\theta}{4}\right).
\end{gather}
From Eq.~(\ref{eq:RzxzExpanded}), we see that $\mathcal{R}_{\rm zxz}\bigr\rvert_{\delta\epsilon_{\rm d} \neq 0}-\mathcal{R}_{\rm zxz}\bigr\rvert_{\delta\epsilon_{\rm d}=0} \sim \left(\delta\epsilon_{\rm d}/g\right)^{2}$, which implies that the error probabilities within the computational subspace scale as $\sim \left(\delta\epsilon_{\rm d}/g\right)^{4}$.
However at this order, the leakage probabilities, $P_{LC}=\left|\bra{L}\mathcal{R}_{\rm zxz}\ket{C}\right|^{2}$ and $P_{LE}=\left|\bra{L}\mathcal{R}_{\rm zxz}\ket{E}\right|^{2} $, both vanish.
Expanding $\mathcal{R}_{\rm zxz}(\theta,\varphi)$ to even higher order in $(\delta\epsilon_\text{d}/g)$ yields
\begin{gather}
P_{LC} = \left(\frac{\delta\epsilon_{\rm d}}{g}\right)^{6}\left|\frac{1}{3}\sin^{2}\left(\frac{\theta}{4}\right)-\frac{\theta}{4}\tan\left(\frac{\theta}{4}\right)+\frac{2}{3}\tan^{2}\left(\frac{\theta}{4}\right)\right|^2 + \mathcal{O}\left(\frac{\delta\epsilon_{\rm d}}{g}\right)^{10}\nonumber \\
P_{LE} = \left(\frac{\delta\epsilon_{\rm d}}{g}\right)^{6}\left|\frac{2}{3}\tan\left(\frac{\theta}{4}\right)-\frac{\theta}{4}+\frac{1}{6}\sin\left(\frac{\theta}{2}\right)\right|^2+ \mathcal{O}\left(\frac{\delta\epsilon_{\rm d}}{g}\right)^{10}.
\end{gather}

\subsection{Analysis of errors for the Identity gate}
\label{sec:identity}
Here we discuss the computational and leakage error probabilities of the Identity gate that we have defined as
\BEq
\label{eq:Identity}
\mathcal{R}_{I}=\left[U_{\rm z}(\epsilon_{\rm q},\delta\epsilon_{\rm d},2\pi)U_{\rm x}(g,\delta\epsilon_{\rm d},2\pi)\right]^2.
\EEq
Taylor expanding $\mathcal{R}_{I}$ about $\delta\epsilon_{\rm d}=0$ gives
\BEq
\label{eq:IdentityExpanded}
\mathcal{R}_{I}=\mathbb{1}_{3\times3}-i\pi\left(1+\frac{4g}{\epsilon_{\rm q}}\right)\left(
\begin{array}{ccc}
 0 & 1 & 0 \\
 1 & 0 & 0 \\
 0 & 0 & 0 \\
\end{array}
\right)\left(\frac{\delta\epsilon_{\rm d}}{g}\right)^2-i\pi\left(1+\frac{4g}{\epsilon_{\rm q}}\right)\left(
\begin{array}{ccc}
 0 & 0 & \frac{-i\pi{g}}{\epsilon_{\rm q}} \\
 0 & 0 & 1 \\
 \frac{-i\pi{g}}{\epsilon_{\rm q}} & 1 & 0 \\
\end{array}
\right)\left(\frac{\delta\epsilon_{\rm d}}{g}\right)^3+ \mathcal{O}\left(\frac{\delta\epsilon_{\rm d}}{g}\right)^{4},
\EEq
where $\mathbb{1}_{3\times3}$ is the $3\times3$ identity matrix. It is interesting to note that no $O[\delta\epsilon_\text{d}/g]$ term occurs in this expansion, due to the special form of Eq.~(\ref{eq:Identity}). 
The $O[(\delta\epsilon_\text{d}/g)^2]$ term cannot be made to vanish because this would require $\epsilon_{\rm q}<0$ (since $g>0$); the condition $\epsilon_{\rm q}<0$ is unphysical here, since it would require a negative gate time.
The resulting computational errors in $\mathcal{R}_{I}$ and $\mathcal{R}_{\rm zxz}$ are therefore both $O[\left(\delta\epsilon_{\rm d}/g\right)^{4}]$. 
For $\mathcal{R}_{I}$, we obtain the following leading-order results for the computational and leakage error probabilities:
\begin{gather}
P_{EC}=\left|\bra{E}\mathcal{R}_{I}\ket{C}\right|^{2} = \left|\bra{C}\mathcal{R}_{I}\ket{E}\right|^{2} = \left(\pi+\frac{4\pi{g}}{\epsilon_{\rm q}}\right)^2\left(\frac{\delta\epsilon_{\rm d}}{g}\right)^4+\mathcal{O}\left(\frac{\delta\epsilon_{\rm d}}{g}\right)^{8}, \nonumber \\
P_{LC}=\left|\bra{L}\mathcal{R}_{I}\ket{C}\right|^{2} = \left(\frac{\pi{g}}{\epsilon_{\rm q}}\right)^{2}\left(\pi+\frac{4\pi{g}}{\epsilon_{\rm q}}\right)^2\left(\frac{\delta\epsilon_{\rm d}}{g}\right)^6 +\mathcal{O}\left(\frac{\delta\epsilon_{\rm d}}{g}\right)^{10} , \nonumber \\
P_{LE}=\left|\bra{L}\mathcal{R}_{I}\ket{E}\right|^{2} = \left(\pi+\frac{4\pi{g}}{\epsilon_{\rm q}}\right)^2\left(\frac{\delta\epsilon_{\rm d}}{g}\right)^6+\mathcal{O}\left(\frac{\delta\epsilon_{\rm d}}{g}\right)^{10}.
\label{eq:PerrorRI}
\end{gather}
Here we note that the limits $\epsilon_\text{q}$ or $g\rightarrow 0$ appear to cause the error probabilities to diverge.  However these limits would require infinite gate times in Eq.~(\ref{eq:Identity}).  Therefore they do not correspond to valid gates.

\subsection{Error function pulse profile}
\label{sec:pulse}

Control pulses from an arbitrary waveform generator (AWG) are smoothed out, intentionally or unintentionally, before they reach the physical qubit inside the dilution refrigerator. 
In our numerical simulations, we model these smooth pulse shapes as error functions~\cite{PhysRevA.87.022309}.
If the initial and final times of the switch are $t_1$ and $t_2$, and the corresponding frequencies are $f_1$ and $f_2$, then we define the switching function as
\BEq
\label{eq:erf}
f(t)=\frac{f_{1}+f_{2}}{2}+\frac{f_{2}-f_{1}}{2}{\rm Erf}\left[\frac{4}{t_{2}-t_{1}}\left(t-\frac{t_{1}+t_{2}}{2}\right)\right],
\EEq
where,
\BEq
{\rm Erf}[z]=\frac{1}{\sqrt{\pi}}\int_{-z}^{z}e^{-k^{2}}dk.
\EEq 
For typical experiments in quantum dots, the switching or rise time is of order $t_2-t_1=50$~ps, which is the value we use in our simulations.

\subsection{Comparison between single and composite pulse approaches}
\label{sec:errorComp}

\begin{figure}[htb]
\centering
\includegraphics[angle=0,width=0.7\linewidth]{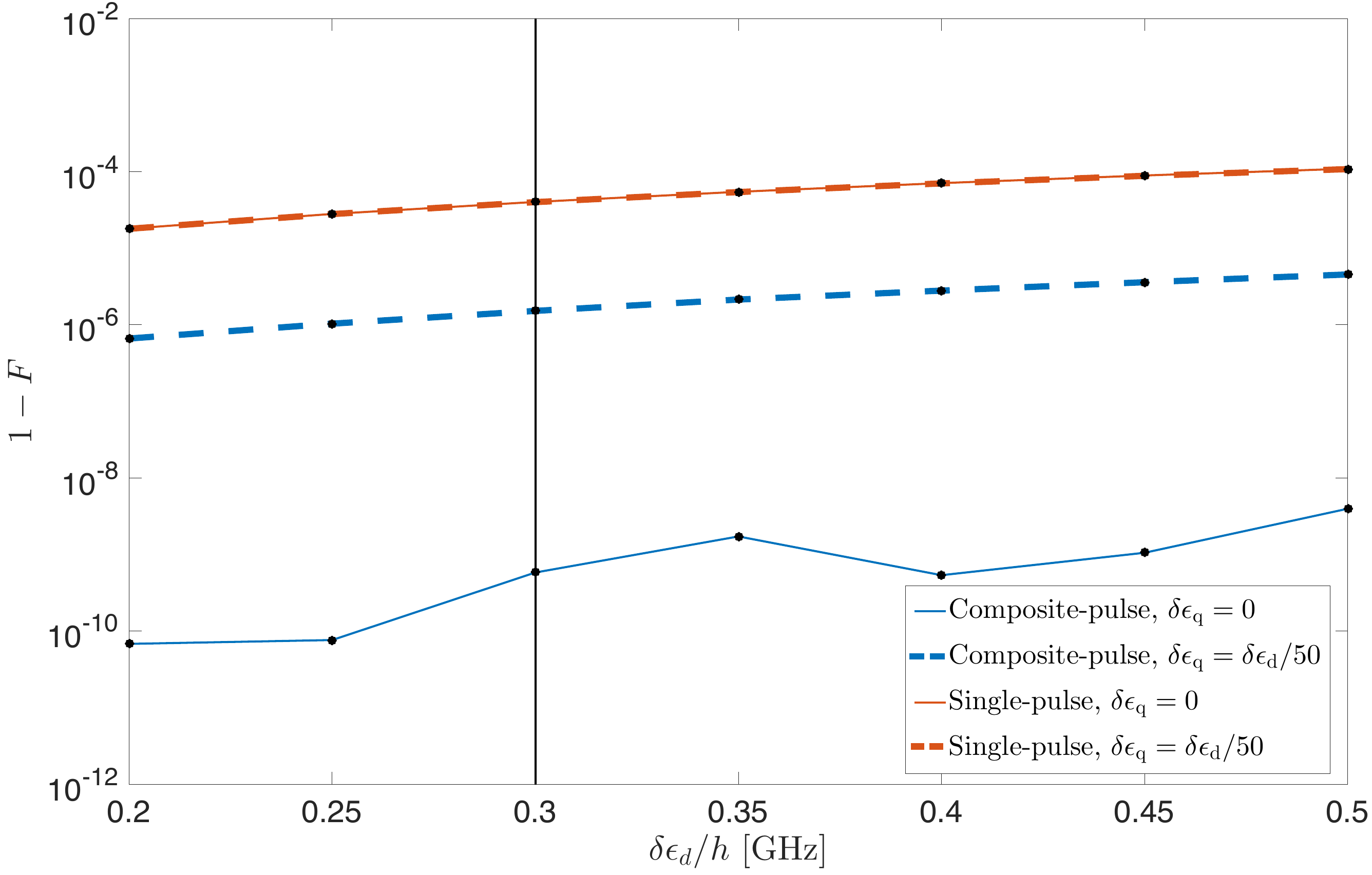}
\caption{{Comparison of the infidelities of $X_{-\pi/2}$ gates as a function of dipolar fluctuation $\delta\epsilon_{\rm d}$ for the composite- and single-pulse approaches. Black-filled circles show the data points that are numerically computed for optimal pulses corresponding to the given $\delta\epsilon_{\rm d}$ and $\delta\epsilon_{\rm q}$. The vertical solid black line indicates a typical value of $\delta\epsilon_{\rm d}$ that is consistent with recent experiments~\cite{Mi_2017}. The dashed lines (note that the red dashed line falls on top of the red solid line) denote infidelity for $\delta\epsilon_{\rm q}=\delta\epsilon_{\rm d}/50$ and the solid lines denote the same for $\delta\epsilon_{\rm q}=0$. The plot demonstrates that the composite pulses yield a large decrease in infidelity over a broad range of amplitudes of the dipolar noise.}}
\label{fig:infidelityComp}
\end{figure}

{Here we compare our composite pulse approach against the single-pulse approach to designing quantum gates for CQ qubit under experimental conditions. In order to obtain a fair comparison, we optimize three pulse parameters for both approaches by varying $\delta\epsilon_{\rm d}$ over a range of experimentally realistic values. Fig.~{\ref{fig:infidelityComp}} shows the results of an infidelity calculation for $X_{-\pi/2}$ gates as a function of $\delta\epsilon_{\rm d}$, based on Eq.~(10) of the main text. These results indicate that in the limit $\delta\epsilon_{\rm q} \rightarrow 0$, the infidelity obtained with our composite pulse sequence is orders of magnitude better than that obtained using an optimized pulse in the single-pulse approach. For $\delta\epsilon_{\rm q} > 0$, we find that the infidelity of the gate operation using the composite pulse sequence is $\delta\epsilon_{\rm q}$-limited, as indicated by the dashed blue line in Fig.~{\ref{fig:infidelityComp}}. However, as discussed in ref.~\cite{Friesen:2016aa}, the ratio $\delta\epsilon_{\rm q}/\delta\epsilon_{\rm d} \approx d/R$ can be very small, where $d$ is the interdot spacing and $R$ is the characteristic distance between the qubit and the charge fluctuators that cause $\delta\epsilon_{\rm d}$. Hence, $\delta\epsilon_{\rm q}$ can be systematically reduced with advanced fabrication techniques that, in turn, reduce the infidelity of quantum gates performed using our approach. This is one of the greatest advantages of using our composite pulse sequence for CQ qubits.}


%

\end{document}